\documentstyle[twoside,epsf]{article}

\catcode`\@=11
\long\def\@makefntext#1{
\protect\noindent \hbox to 3.2pt {\hskip-.9pt
$^{{\eightrm\@thefnmark}}$\hfil}#1\hfill}       

\def\@makefnmark{\hbox to 0pt{$^{\@thefnmark}$\hss}}    

\def\ps@myheadings{\let\@mkboth\@gobbletwo
\def\@oddhead{\hbox{}
\rightmark\hfil\eightrm\thepage}
\def\@oddfoot{}\def\@evenhead{\eightrm\thepage\hfil
\leftmark\hbox{}}\def\@evenfoot{}
\def\sectionmark##1{}\def\subsectionmark##1{}}

\oddsidemargin=\evensidemargin
\newcounter{sectionc}\newcounter{subsectionc}\newcounter{subsubsectionc}
\renewcommand{\section}[1] {\vspace{12pt}\addtocounter{sectionc}{1}
\setcounter{subsectionc}{0}\setcounter{subsubsectionc}{0}\noindent
        {\tenbf\thesectionc. #1}\par\vspace{5pt}}
\renewcommand{\subsection}[1] {\vspace{12pt}\addtocounter{subsectionc}{1}
        \setcounter{subsubsectionc}{0}\noindent
        {\bf\thesectionc.\thesubsectionc. {\kern1pt \bfit #1}}\par\vspace{5pt}}
\renewcommand{\subsubsection}[1] {\vspace{12pt}\addtocounter{subsubsectionc}{1}
        \noindent{\tenrm\thesectionc.\thesubsectionc.\thesubsubsectionc.
        {\kern1pt \tenit #1}}\par\vspace{5pt}}
\newcommand{\nonumsection}[1] {\vspace{12pt}\noindent{\tenbf #1}
        \par\vspace{5pt}}

\newcounter{appendixc}
\newcounter{subappendixc}[appendixc]
\newcounter{subsubappendixc}[subappendixc]
\renewcommand{\thesubappendixc}{\Alph{appendixc}.\arabic{subappendixc}}
\renewcommand{\thesubsubappendixc}
    {\Alph{appendixc}.\arabic{subappendixc}.\arabic{subsubappendixc}}

\renewcommand{\appendix}[1] {\vspace{12pt}
        \refstepcounter{appendixc}
        \setcounter{figure}{0}
        \setcounter{table}{0}
        \setcounter{lemma}{0}
        \setcounter{theorem}{0}
        \setcounter{corollary}{0}
        \setcounter{definition}{0}
        \setcounter{equation}{0}
        \renewcommand{\thefigure}{\Alph{appendixc}.\arabic{figure}}
        \renewcommand{\thetable}{\Alph{appendixc}.\arabic{table}}
        \renewcommand{\theappendixc}{\Alph{appendixc}}
        \renewcommand{\thelemma}{\Alph{appendixc}.\arabic{lemma}}
        \renewcommand{\thetheorem}{\Alph{appendixc}.\arabic{theorem}}
        \renewcommand{\thedefinition}{\Alph{appendixc}.\arabic{definition}}
        \renewcommand{\thecorollary}{\Alph{appendixc}.\arabic{corollary}}
        \renewcommand{\theequation}{\Alph{appendixc}.\arabic{equation}}
        \noindent{\tenbf Appendix \theappendixc #1}\par\vspace{5pt}}
\newcommand{\subappendix}[1] {\vspace{12pt}
        \refstepcounter{subappendixc}
        \noindent{\bf Appendix \thesubappendixc. {\kern1pt \bfit #1}}
    \par\vspace{5pt}}
\newcommand{\subsubappendix}[1] {\vspace{12pt}
        \refstepcounter{subsubappendixc}
        \noindent{\rm Appendix \thesubsubappendixc. {\kern1pt \tenit #1}}
    \par\vspace{5pt}}

\newcommand{\textlineskip}{\baselineskip=13pt}
\newcommand{\smalllineskip}{\baselineskip=10pt}

\def\eightcirc{
\begin{picture}(0,0)
\put(4.4,1.8){\circle{6.5}}
\end{picture}}
\def\eightcopyright{\eightcirc\kern2.7pt\hbox{\eightrm c}}

\newcommand{\copyrightheading}[1]
        {\vspace*{-2.5cm}\smalllineskip{\flushleft
        {\footnotesize International Journal of Modern Physics C, #1}\\
        {\footnotesize $\eightcopyright$\,\,\, World Scientific Publishing
         Company}\\
         }}

\newcommand{\publisher}[2]{{\begin{center}\footnotesize\smalllineskip
        Received #1\\
        Revised #2
        \end{center}
        }}

\def\abstracts#1#2#3{{
        \centering{\begin{minipage}{4.5in}\baselineskip=10pt\footnotesize
        \parindent=0pt #1\par
        \parindent=15pt #2\par
        \parindent=15pt #3\par
        \end{minipage}}\par}}

\newcommand{\bibit}{\nineit}
\newcommand{\bibbf}{\ninebf}
\renewenvironment{thebibliography}[1]
        {\frenchspacing
     \ninerm\baselineskip=11pt
         \begin{list}{\arabic{enumi}.}
        {\usecounter{enumi}\setlength{\parsep}{0pt}
         \setlength{\leftmargin 17pt}{\rightmargin 0pt}   
         \setlength{\itemsep}{0pt} \settowidth
    {\labelwidth}{#1.}\sloppy}}{\end{list}}

\newcounter{itemlistc}
\newcounter{romanlistc}
\newcounter{alphlistc}
\newcounter{arabiclistc}

\newcommand{\fcaption}[1]{
        \refstepcounter{figure}
    \setbox\@tempboxa = \hbox{\footnotesize Fig.~\thefigure. #1}
    \ifdim \wd\@tempboxa > 5in
           {\begin{center}
    \parbox{5in}{\footnotesize\smalllineskip Fig.~\thefigure. #1}
            \end{center}}
        \else
             {\begin{center}
         {\footnotesize Fig.~\thefigure. #1}
              \end{center}}
        \fi}

\newcommand{\tcaption}[1]{
        \refstepcounter{table}
    \setbox\@tempboxa = \hbox{\footnotesize Table~\thetable. #1}
        \ifdim \wd\@tempboxa > 5in
           {\begin{center}
         \parbox{5in}{\footnotesize\smalllineskip Table~\thetable. #1}
            \end{center}}
        \else
             {\begin{center}
         {\footnotesize Table~\thetable. #1}
              \end{center}}
        \fi}

\def\@citex[#1]#2{\if@filesw\immediate\write\@auxout
    {\string\citation{#2}}\fi
\def\@citea{}\@cite{\@for\@citeb:=#2\do
    {\@citea\def\@citea{,}\@ifundefined
    {b@\@citeb}{{\bf ?}\@warning
    {Citation `\@citeb' on page \thepage \space undefined}}
    {\csname b@\@citeb\endcsname}}}{#1}}

\newif\if@cghi
\def\cite{\@cghitrue\@ifnextchar [{\@tempswatrue
    \@citex}{\@tempswafalse\@citex[]}}
\def\citelow{\@cghifalse\@ifnextchar [{\@tempswatrue
    \@citex}{\@tempswafalse\@citex[]}}
\def\@cite#1#2{{$\null^{#1}$\if@tempswa\typeout
    {IJCGA warning: optional citation argument
    ignored: `#2'} \fi}}

\def\pmb#1{\setbox0=\hbox{#1}
        \kern-.025em\copy0\kern-\wd0
        \kern.05em\copy0\kern-\wd0
        \kern-.025em\raise.0433em\box0}

\def\fnt#1#2{\footnotetext{\kern-.3em
        {$^{\mbox{\scriptsize #1}}$}{#2}}}

\def\fpage#1{\begingroup
\voffset=.3in
\thispagestyle{empty}\begin{table}[b]\centerline{\footnotesize #1}
        \end{table}\endgroup}

\def\runninghead#1#2{\pagestyle{myheadings}
\markboth{{\protect\footnotesize\it{\quad #1}}\hfill}
{\hfill{\protect\footnotesize\it{#2\quad}}}}
\font\tenbf=cmbx10
\font\tenit=cmti10
\font\tenit=cmti10
\font\bfit=cmbxti10 at 10pt
\font\ninebf=cmbx9
\font\ninerm=cmr9
\font\nineit=cmti9

\font\eightrm=cmr8

\def\lsym{\raise-3pt\hbox{\vbox{\tabskip0pt\offinterlineskip
    \halign{\tabskip0pt plus 1em
    ##\tabskip0pt\cr
    $\,\,<\,\,$\cr
    $\,\,\sim\,\,$\cr}}}}
\def\rsym{\raise-3pt\hbox{\vbox{\tabskip0pt\offinterlineskip
     \halign{\tabskip0pt plus 1em
      ##\tabskip0pt\cr
      $\,\,>\,\,$\cr
      $\,\,\sim\,\,$\cr}}}}
\def\qed{\hbox{${\vcenter{\vbox{            
    \hrule height 0.4pt\hbox{\vrule width 0.4pt height 6pt
    \kern5pt\vrule width 0.4pt}\hrule height 0.4pt}}}$}}
\def\theequation{\thesection.\arabic{equation}}     

\newcommand{\ZZ}{{\mathbf{Z}}}
\newcommand{\NN}{{\mathbf{N}}}

\begin{document}

\runninghead{N. Boccara \& M. Roger}
{Periodic behavior in 2D  Cellular Automata}

\normalsize\textlineskip
\thispagestyle{empty}
\setcounter{page}{1}

\copyrightheading{Vol. 0, No. 0 (1993) 000--000}

\vspace*{0.88truein}

\fpage{1}
\centerline{\bf }
\vspace*{0.035truein}
\centerline{\bf TOTALISTIC TWO-DIMENSIONAL CELLULAR AUTOMATA}
\vspace*{0.035truein}
\centerline{\bf EXHIBITING GLOBAL PERIODIC BEHAVIOR}
\vspace*{0.37truein}
\centerline{\footnotesize N. BOCCARA\footnote{Permanent address: University
of Illinois at Chicago, Dept. of Physics, Chicago, IL 60607-7059,
USA}~\ and M. ROGER}
\vspace*{0.015truein}
\centerline{\footnotesize\it DRECAM SPEC Centre d'\'Etudes de Saclay,}
\baselineskip=10pt
\centerline{\footnotesize\it 91191 Gif-sur-Yvette Cedex, France}
\vspace*{0.15truein}

\vspace*{0.225truein}
\publisher{  }{  }

\vspace*{0.21truein} \abstracts{We have determined families of
two-dimensional deterministic totalistic cellular automaton rules
whose stationary density of active sites exhibits a period two in
time. Each family of deterministic rules is characterized by an
``average probabilistic totalistic rule'' exhibiting the
same periodic behavior.}{}{}

\vspace*{0.2in}

Many natural populations of plants and animals exhibit large
fluctuations of density with a roughly cyclic behavior. A
well-known example is the oscillatory behavior of the Canadian
lynx population as documented in the data compiled by the Hudson
Bay Company over the period 1735-1940. Oscillations with an
approximate period of about 10 years are observed with large
amplitude fluctuations, which could, actually, correspond to a
chaotic behavior.\cite{schaf1,schaf2} Good introductions to
population dynamics may be found in May\cite{may} and
Murray.\cite{mur} Most models in population dynamics are formulated
in terms of differential equations or difference equations, which
means that the local character of the interactions between prey
and predators, for example, is not taken into account. In order to
describe more correctly the local character of the predation process,
it would be better to formulate predator-prey models in terms of cellular
automata (CAs), which are fully discrete dynamical systems
evolving in time according to local rules. This paper does not
deal with any specific model. Its purpose is to understand under
which circumstances CAs exhibit nontrivial global behaviors. It
has been argued\cite{grin} that, in spatially isotropic systems
with short-range interactions evolving in discrete time, periodic
states with periods larger than 2, are never stable under generic
conditions. Despite these arguments, Chat\'e and
Manneville\cite{chat1,chat2} gave several examples of
high-dimensional deterministic CAs exhibiting a global
(quasi)period-3 behavior. In order to verify that the
corresponding states were not metastable Gallas {\it et
al.}\cite{gall} performed a detailed study of the
Chat\'e--Manneville CAs, and confirmed that the nontrivial
behavior found by these authors extends to much longer
times and to much larger lattices. Since the chance to find
such a behavior should increase with the dimension of the
CA, it is of interest to find low-dimensional CAs exhibiting the
same behavior. A three-dimensional deterministic CA exhibiting a
quasiperiodic behavior has been found by Hemmingson\cite{hem},
and shortly after its nontrivial collective behavior has been
confirmed through large-scale simulations (Chat\'e {\it et
al.}\cite{chat3}).

In this note we present families of two-dimensional totalistic
class-3 CAs which exhibit a period-2 behavior. The class-3
requirement eliminates CAs whose attractor consists of only two
points in the configuration space. No arguments have been given
against such a collective behavior, but these systems might be of
interest since they are not so frequent. Moreover, for each
neighborhood defining a CA family, there exists a
well-defined ``average probabilistic CA rule'' which exhibits
the same collective behavior.

Two-dimensional deterministic totalistic CAs may be defined as follows:
Let $s:\ZZ\times\ZZ\times\NN\mapsto\{0,1\}$ be a function that
satisfies the equation
$$
s(i_1,i_2,t+1)=f\Big(\sum_{\{j_1,j_2\}\in {\cal V}_r}
s(i_1+j_1,i_2+j_2,t)\Big),
$$
for all $(i_1,i_2)\in\ZZ^2$ and all $t\in\NN$, where $\ZZ$ is the set of all
integers and $\NN$ the set of nonnegative integers.
${\cal V}_r$ defines a neighborhood of site $(i_1,i_2)$ which is
either of the ``von~Neuman-type'' with
$$
|j_1|+|j_2|\le r
$$
or of the ``Moore-type'' with
$$
|j_1|\le r \qquad {\rm and} \qquad |j_2|\le r
$$
The integer $r$ characterizes the range of the neighborhood. For a given
neighborhood, each totalistic rule is, therefore, determined by a function
$$
S\mapsto f(S),
$$
where
$$
S\in\{0,1,2,\ldots,S_{\max}\}\qquad {\rm and} \qquad f(S)\in\{0,1\},
$$
with $S_{\max}=(2r+1)^2$ for a Moore-type neighborhood and
$S_{\max}=2r^2+2r+1$ for a von~Neuman-type neighborhood.

For a given neighborhood, the number of rules is $2^{S_{\max}}$. For relatively
small neighborhoods $S_{\max}\le 13$, a systematic search for rules exhibiting
nontrivial collective behavior is possible. For the simplest Moore and
von Neuman neighborhoods, with $r=1$ we did not find any nontrivial behavior.
The smallest neighborhood for which such a behavior has been
found is the von~Neuman neighborhood with $r=2$, and the corresponding
family of $2^{14}=16384$ rules has been systematically investigated. Systematic
investigations of families of CA rules with range larger than 2 is impossible. In
this case, for each family, we studied a randomly selected sample of about
50000 rules.

\vspace{0.2cm}

\noindent\textbf{von Neuman neighborhood with $r=2$.}

\vspace{0.2cm}

\noindent Each rule has been first studied on a $128\times 128$-lattice over 200
time steps. The average concentrations $c_e$ and $c_o$ of nonzero sites over
the 50 last even and odd time steps were evaluated. We only retained rules
satisfying the condition $|c_e-c_o|>0.1$. A few hundred rules satisfying
this criterion have been thoroughly investigated on a $512\times 512$-lattice
over 20000 time steps . Several behaviors have been observed.

\begin{itemize}
\item Some of them are ``similar'' to Conway's \textit{Game of life}, in the sense
that only periodic small structures, with various periods, remain on a
period-2 background (see Figure 1). In this case, the attractor consists of a very
few number of points in the configuration space.

\begin{figure}[htbp]
\epsfxsize=7truecm
\epsfbox{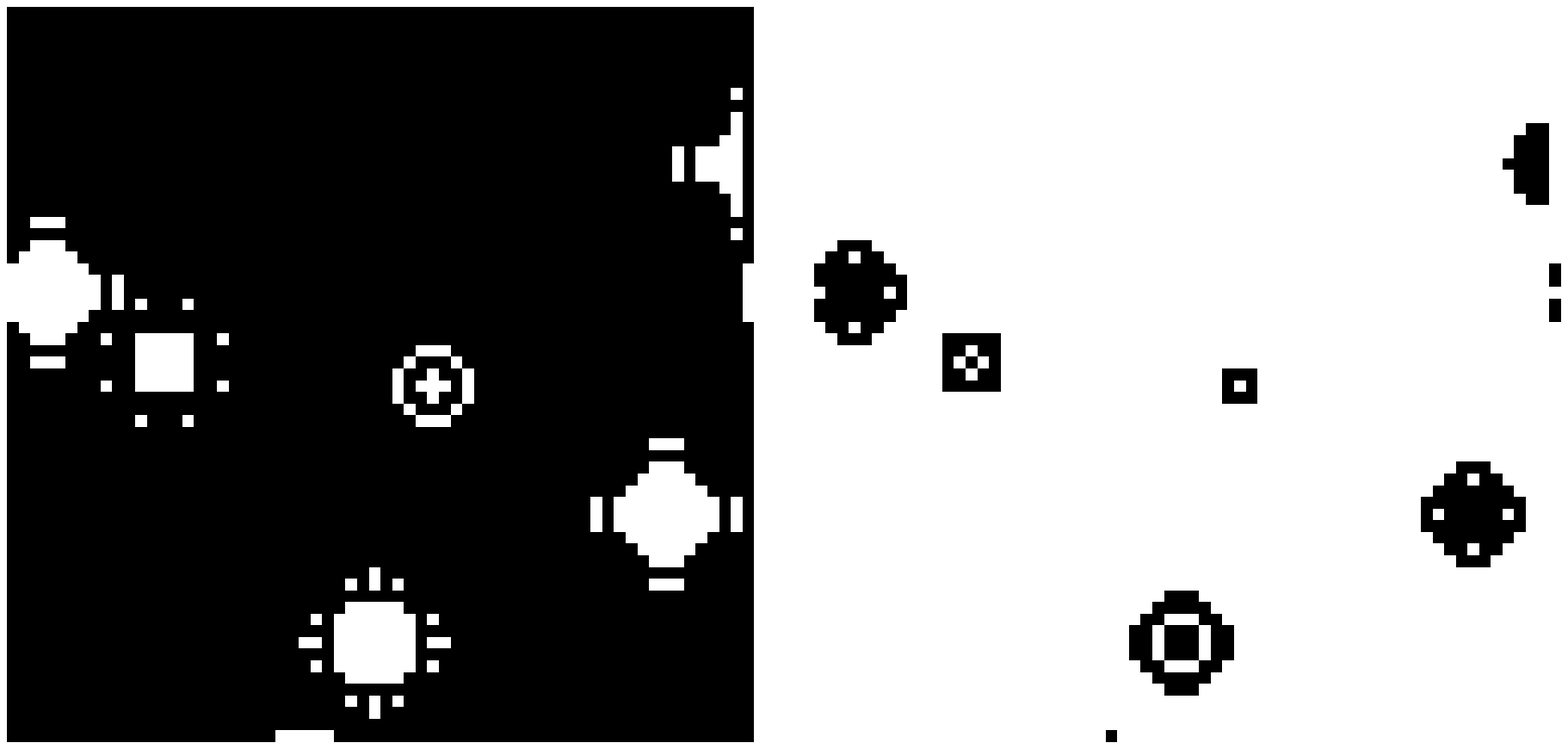}
\fcaption{Rule 189: periodic structures on a period-2 background.}
\end{figure}

\item Some converge to a fixed number of nonzero sites after large transient
time during which the concentration oscillates. Their patterns are inhomogeneous
(see Figure 2). The ``local'' concentration of nonzero sites oscillates;
however, at large times, in the infinite-size limit, the concentration of nonzero
sites tends to a fixed point (see Figure 2). Such a behavior has already been
observed in some probabilistic cellular automata\cite{roblin}.

\begin{figure}[htbp]
\epsfxsize=7truecm
\epsfbox{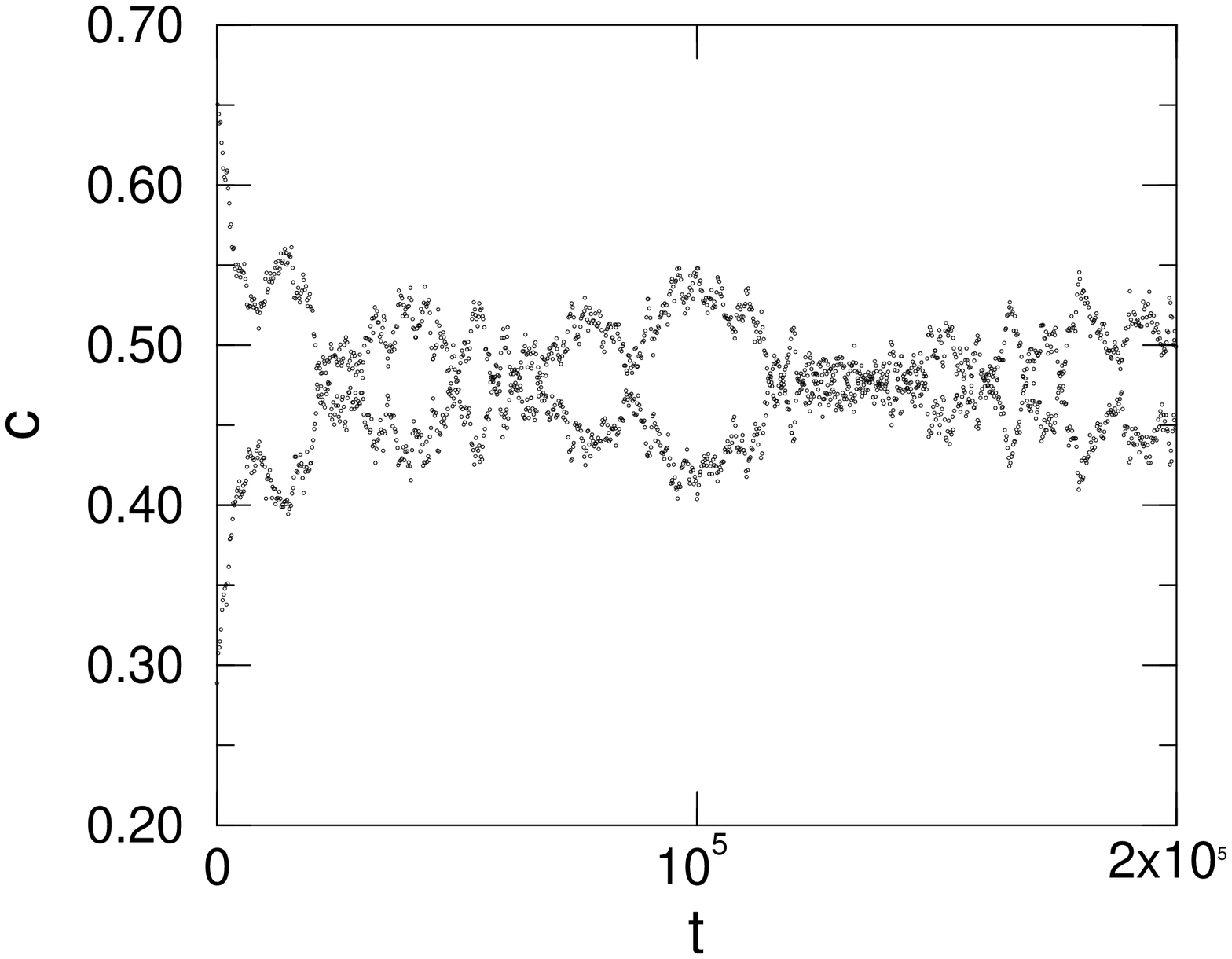}
\fcaption{Rule 2158: evolution of the concentration $c$ of nonzero sites on
a $512\times 512$-lattice.}
\end{figure}

\begin{figure}[htbp]
\epsfxsize=7truecm
\epsfbox{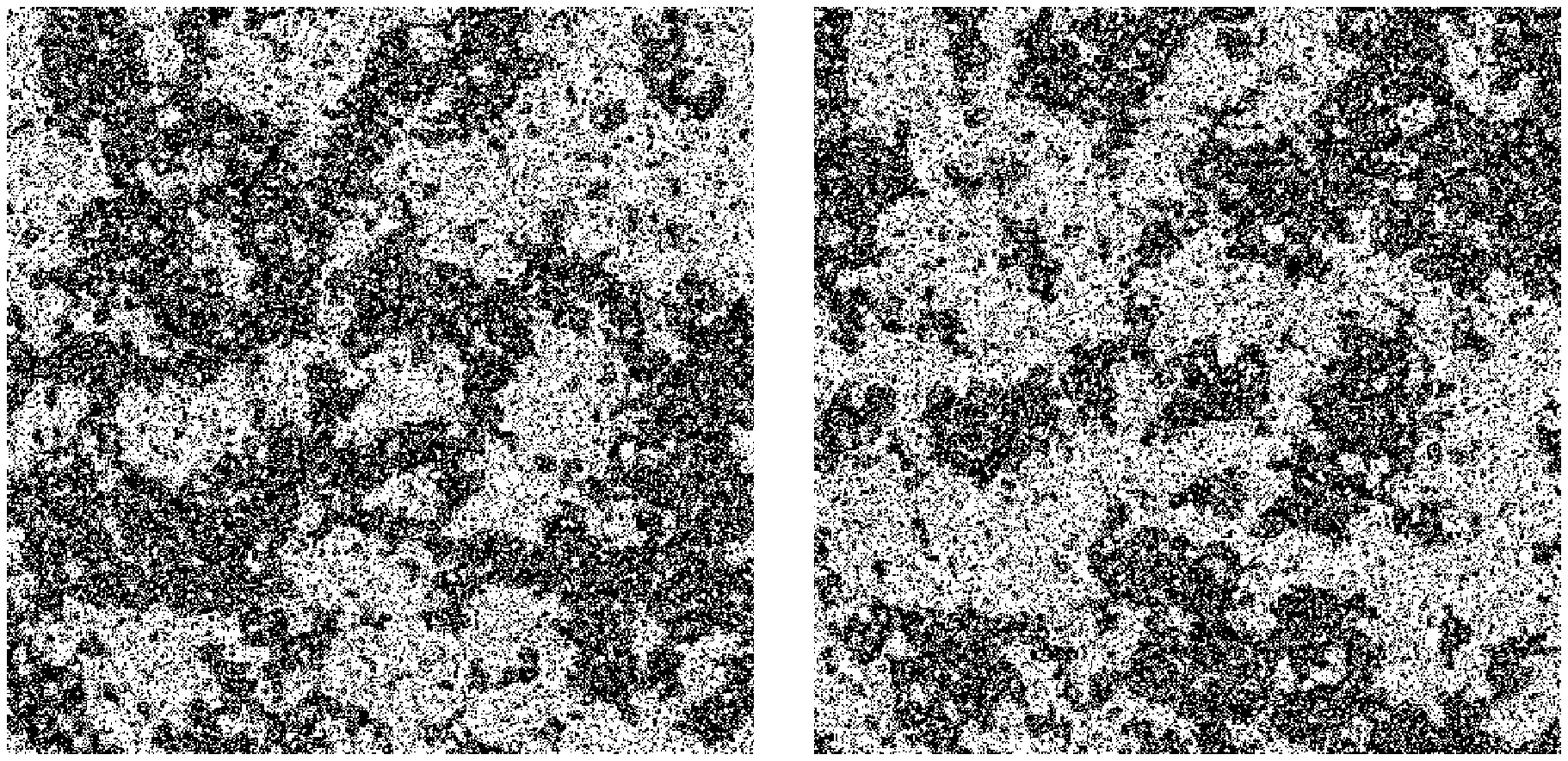}
\fcaption{Spatial patterns of Rule 2158 at two consecutive time steps after
$2\times 10^4$ iterations. The lattice has $512\times 512$ sites. There are
short range correlations over a length scale which is much larger than the
range $r=2$ of the rule. However the concentration of nonzero sites tends
to a fixed point on an infinite lattice.}
\end{figure}

\item The 54 rules listed in Table I have a nontrivial collective period-2
behavior for the concentration of nonzero sites. They appear by pairs,
rule $g$ defined by
$$
g(S)=1-f(S_{max}-S)
$$
has the same properties as $f$. A typical example (Rule 2077) is shown
in Figures~4 and 5. Note that all these rules are such that $f(0)=1$.
\end{itemize}

\begin{figure}[htbp]
\epsfxsize=7truecm
\epsfbox{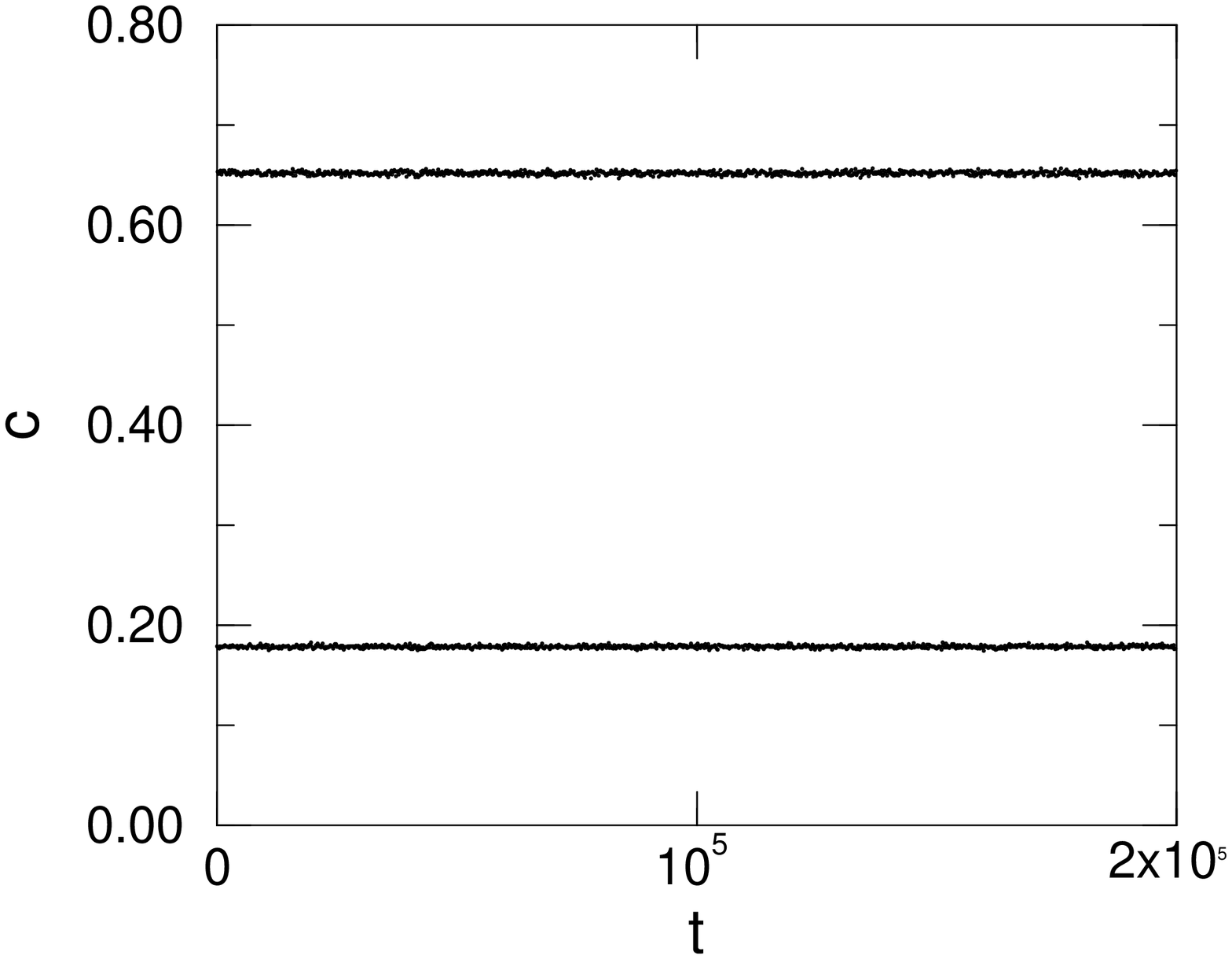}
\fcaption{Rule 2077: evolution of the concentration of nonzero sites as
a function of time on a $512\times 512$-lattice. A period-2 behavior is
clearly observed.}
\end{figure}

\begin{figure}[htbp]
\epsfxsize=7truecm
\epsfbox{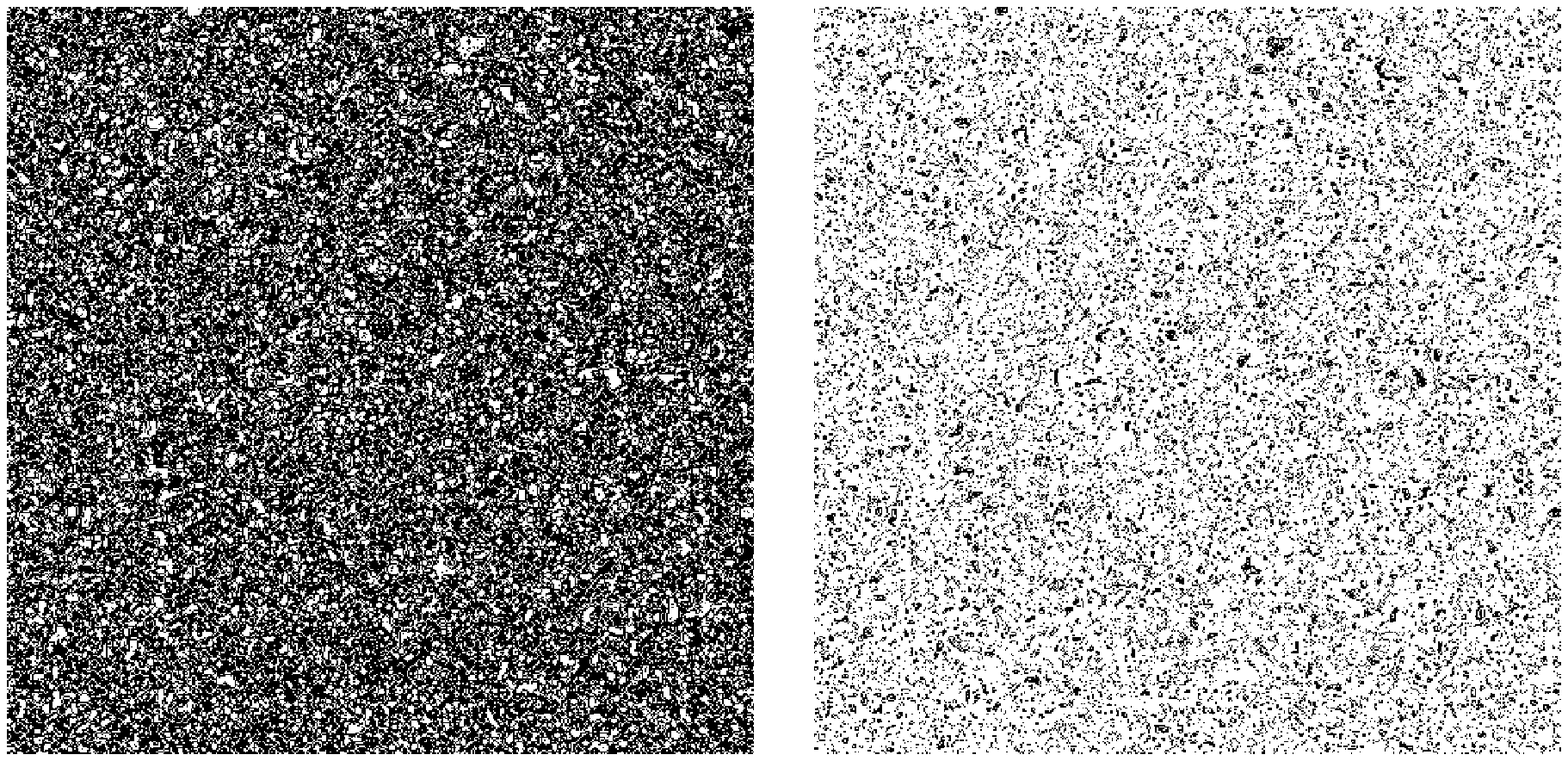}
\fcaption{Patterns of the Rule 2077 at two consecutive times after
$2\times 10^4$ iterations.}
\end{figure}

\begin{table}[htbp]
\tcaption{Deterministic totalistic CA rules exhibiting period-2
collective behavior with a range-2 von-Neuman neighborhood.}
\centerline{\footnotesize}
\centerline{\footnotesize\smalllineskip
\begin{tabular}{l l l l}\\
\hline
{}00000100111101 & 317 &   01000011011111 & 4319\\
{}00000110111101 & 445 &   01000010011111 & 4255\\
{}00001000111101 & 573 &   01000011101111 & 4335\\
{}00001001111101 & 637 &   01000001101111 & 4207\\
{}00010000110111 & 1079 &  00010011110111 & 1271\\
{}00010000111011 & 1083 &  00100011110111 & 2295\\
{}00010000111101 & 1085 &  01000011110111 & 4343\\
{}00010001110111 & 1143\\
{}00010001111011 & 1147\\
{}00010001111101 & 1149 &  01000001110111 & 4215\\
{}00010011111011 & 1275 &  00100000110111 & 2103\\
{}00010011111101 & 1277 &  01000000110111 & 4151\\
{}00100000001101 & 2061 &  01001111111011 & 5115\\
{}00100000011011 & 2075 &  00100111111011 & 2555\\
{}00100000011101 & 2077 &  01000111111011 & 4603\\
{}00100000111011 & 2107 &  00100011111011 & 2299\\
{}00100000111101 & 2109 &  01000011111011 & 4347\\
{}00100001110111 & 2167\\
{}00100001111011 & 2171\\
{}00100001111101 & 2173 &  01000001111011 & 4219\\
{}00100010111101 & 2237 &  01000010111011 & 4283\\
{}00100011111101 & 2301 &  01000000111011 & 4155\\
{}00100101111101 & 2429 &  01000001011011 & 4187\\
{}00100111111101 & 2557 &  01000000011011 & 4123\\
{}01000000001101 & 4109 &  01001111111101 & 5117\\
{}01000000011101 & 4125 &  01000111111101 & 4605\\
{}01000000111101 & 4157 &  01000011111101 & 4349\\
{}01000001011101 & 4189 &  01000101111101 & 4477\\
{}01000001111101 & 4221 &  01000010111101 & 4285\\
\hline\\
\end{tabular}}
\end{table}

In order to characterize a family of rules $\{f_k\mid k\in K\}$,
which exhibit a period two in the infinite-time limit, we define the function
$f_{\rm average}:[0,1]\mapsto[0,1]$ by
$$
f_{\rm average}(S/S_{max})=
\frac{1}{|K|}\sum_{k\in K}f_k(S),
$$
where $|K|$ denotes the number of rules belonging to the $K$ family (see Figure 6).

\begin{figure}[htbp]
\epsfxsize=7truecm
\epsfbox{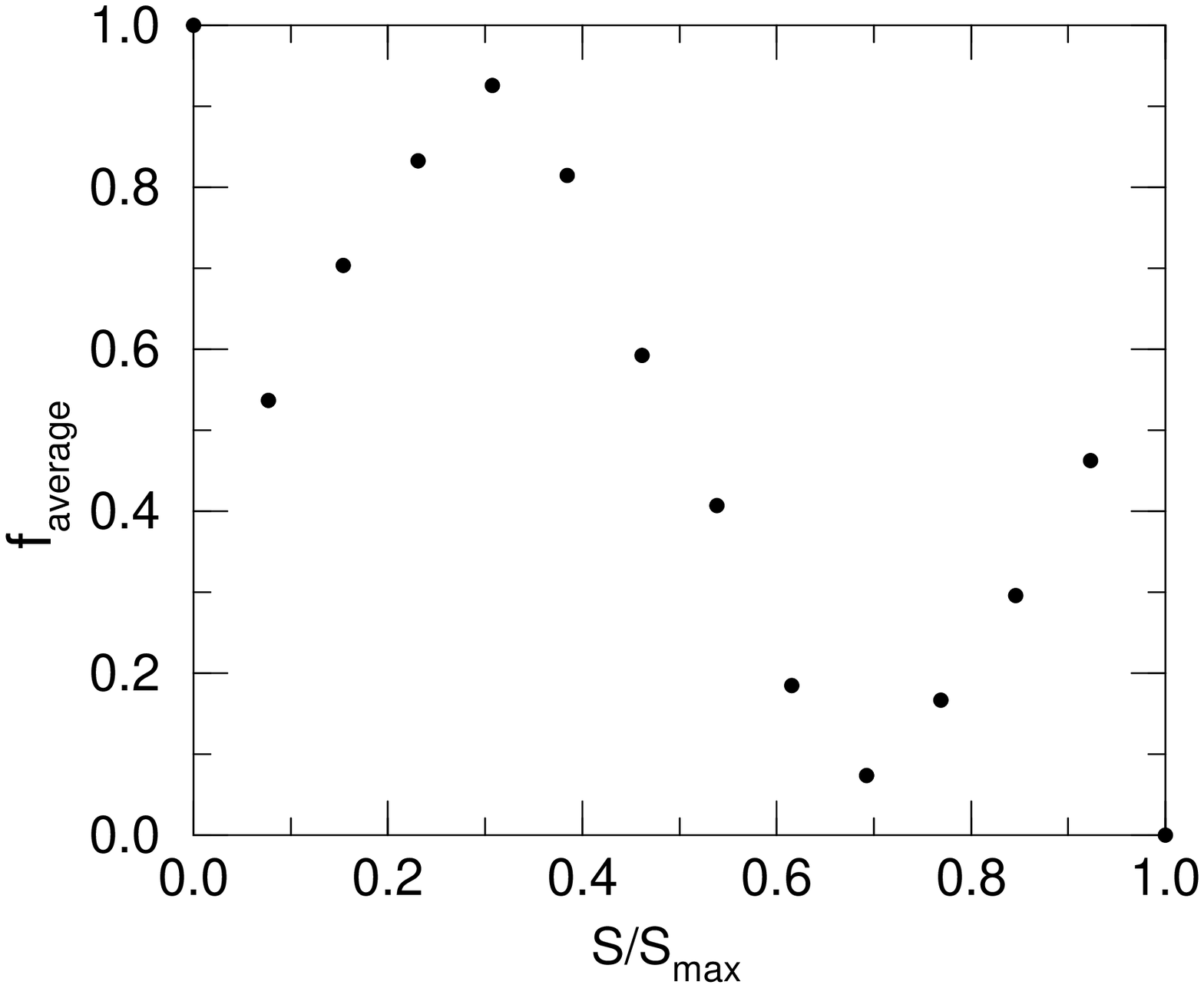}
\fcaption{Graph of the function $f_{\rm average}$ characterizing the
family of rules
defined by a range-2 von Neuman neighborhood.}
\end{figure}

\vspace{0.2cm}

\noindent\textbf{Rules with larger neighborhoods.}

\vspace{0.2cm}

We have studied three other families characterized by the following
neighborhoods.
\begin{itemize}
\item Range-3 von Neuman neighborhood: $S_{\max}=25$, \textit{i.e.\/}
$2^{26}=67108864$ rules.
\item Range-2 Moore neighborhood: $S_{\max}=25$, \textit{i.e.\/}
$2^{26}=67108864$ rules.
\item Range-4 Moore neighborhood: $S_{\max}=81$, \textit{i.e.\/}
$2^{82}\approx 4.836\,10^{24}$ rules.
\end{itemize}

As mentioned above, a systematic investigation of such a large number
of rules is impossible. Therefore, for each family, we only studied a
sample of about 50000 rules selected
at random. We found that a small proportion of these samples
did exhibit a period-2 behavior. \textit{i.e.\/} about 0.7\% of
the rules with a neighborhood containing 25 sites, and 1\% of the
rules with a neighborhood containing 81 sites. In contrast with
the family of rules with a range-2 von Neuman neighborhood,
about 40 to 50\% of them were such that $f(0)=0$. For these
three families, we determined the functions $f_{\rm average}$
whose graphs are represented in Figure~7. These functions are well
approximated by a simple combination of two sine functions of the
form:
$$
f_{\rm average}(x)=0.5-A\sin(2\pi(x-0.5))-B\sin(2\pi(x-0.5))
$$
where $x=S/S_{\max}$. Fitting the data, we found:
\begin{itemize}
\item for the range-3 von Neuman neighborhood, $A\approx 0.291$,
$B\approx 0.008$;
\item for the range-2 Moore neighborhood, $A\approx 0.304$, $B\approx 0.003$;
\item for the range-4 Moore neighborhood, $A\approx 0.138$, $B\approx 0.074$.
\end{itemize}

\begin{figure}[htbp]
\epsfxsize=7truecm
\epsfbox{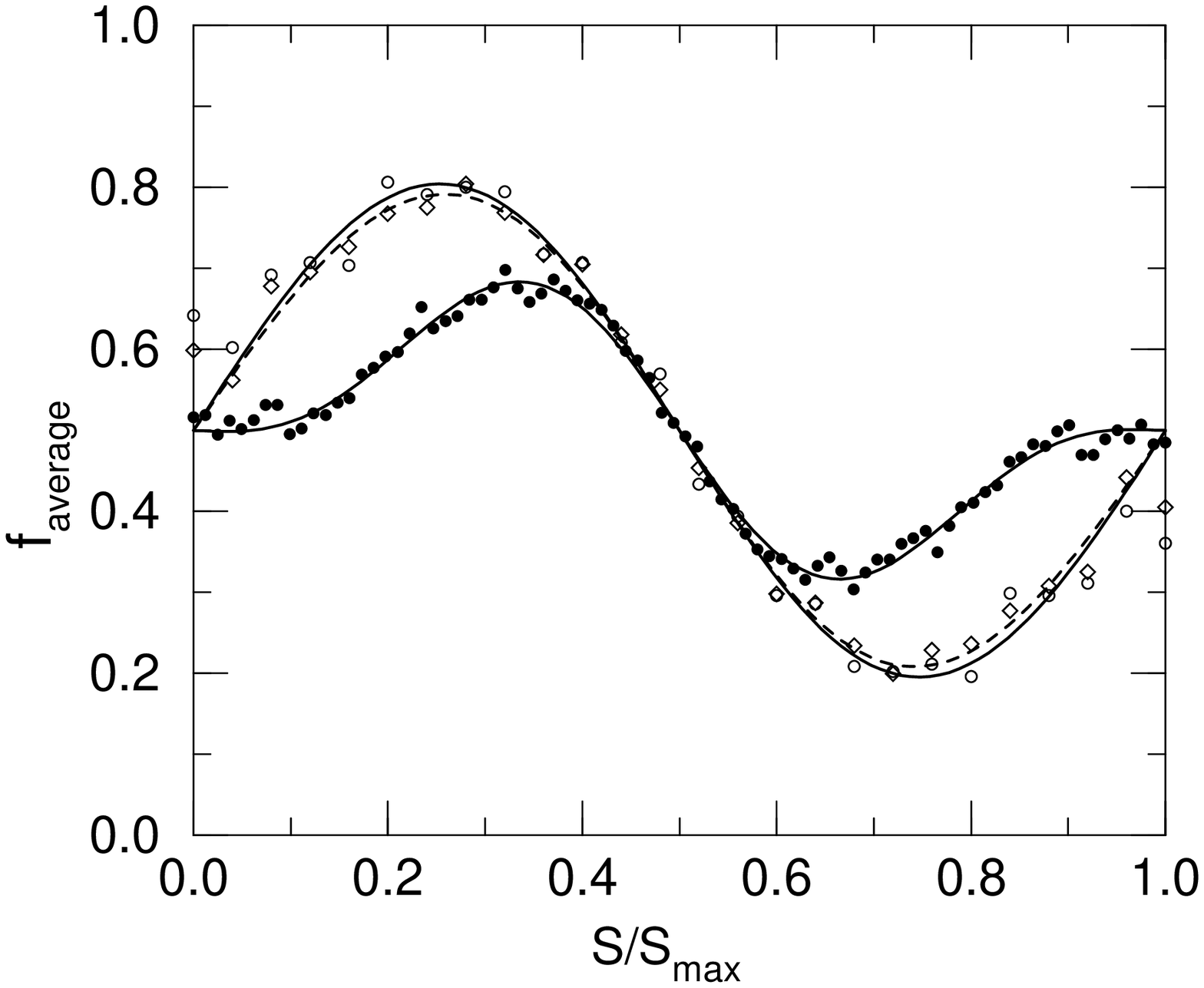}
\fcaption{Graph of functions $f_{\rm average}$ characterizing families of rules
defined by:
(a) range-3 von Neuman neighborhood (diamonds and dashed line),
(b) range-2 Moore neighborhood (open circles and solid line), and
(c) range-4 Moore neighborhood (bullets and solid line).}
\end{figure}

To a function $f_{\rm average}$ we associate a probabilistic totalistic
law defined by
$$
f(S)=
\cases{
 1, & \textrm{with probability $p= f_{average}(S/S_{\max})$;} \cr
 0, & \textrm{with probability $1-p$.}\cr}
$$
Probabilistic totalistic CAs evolving in time according to rules of this
type exhibit period-2 behaviors. As an example, the evolution of the
concentration $c$ of nonzero sites for the probabilistic rule
associated to the family defined by the range-4 Moore neighborhood
is represented in Figure 8. The concentration oscillates between
$c_1=0.337 $ and $c_2=0.663$. The mean-field approximation
predicts the amplitude of the oscillations with a very good
accuracy. Within this approximation, it is found that the
concentration oscillates between $c_1^{\rm mfa}=0.335$
and $c_2^{\rm mfa}=0.665$.

\begin{figure}[htbp]
\epsfxsize=7truecm
\epsfbox{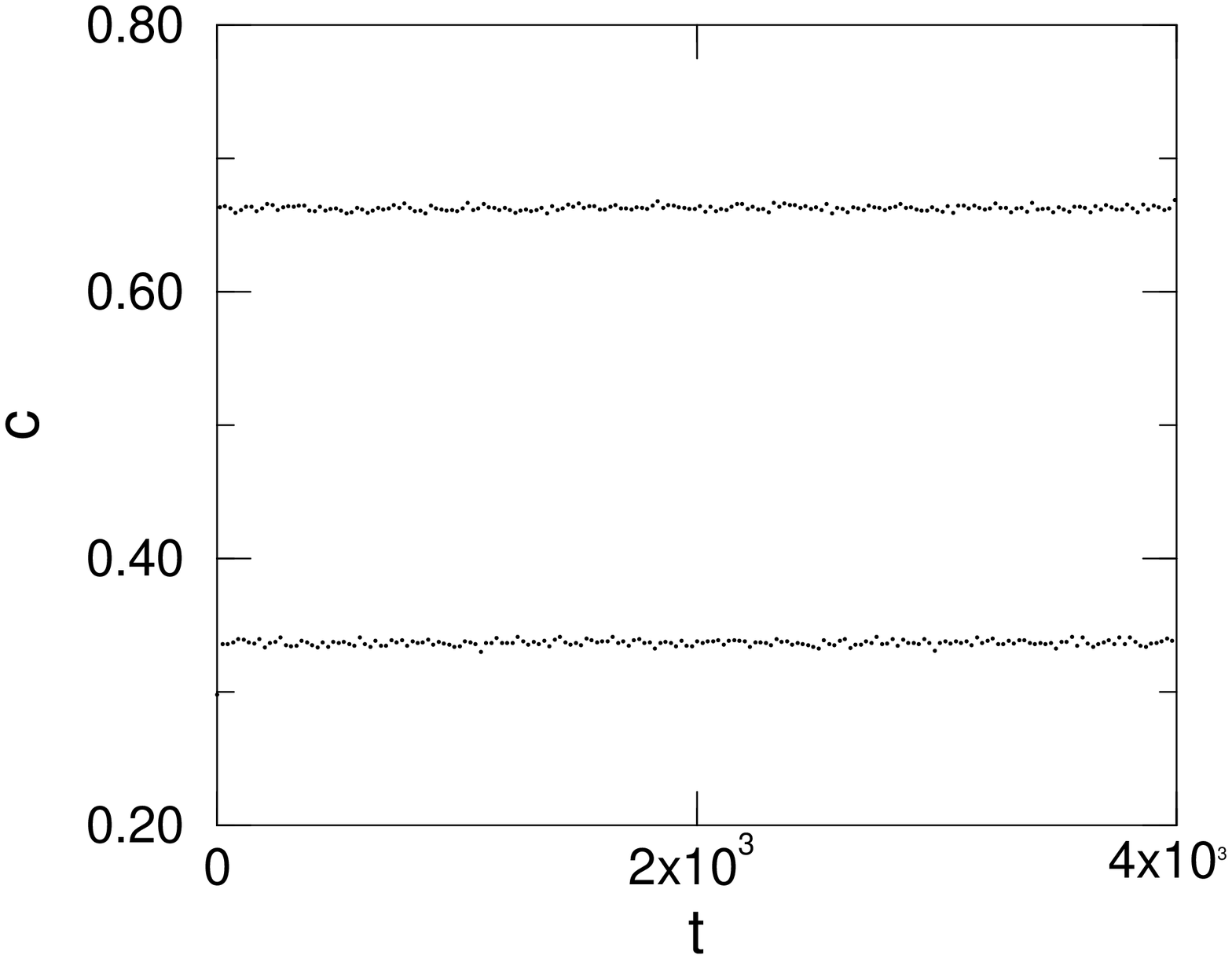}
\fcaption{Concentration of nonzero sites of a probabilistic totalistic CA
whose rule is determined by the function $f_{\rm average}$ characterizing
the family of rules defined by the range-4 Moore neighborhood.}
\end{figure}

The proportion of deterministic totalistic class-3 CAs whose density of nonzero
sites is periodic in time increases with the range of the neighborhood
characterizing the evolution rule. In agreement with an argument due to
Grinstein,\cite{grin} no period larger than 2 has been observed.
To each family of rules exhibiting a period-2 behavior, it is
possible to associate a well-defined mapping from $[0,1]$ to
$[0,1]$. Such a mapping defines a probabilistic totalistic CA
rule, and the density of nonzero sites of a CA evolving according
to such a rule also exhibits a period-2 behavior. For these
probabilistic CAs, the mean-field approximation predicts the values
of the density with a good accuracy.

\nonumsection{References}

\end{document}